# Smart Households Demand Response Management with Micro Grid

Hossein Mohammadi Ruzbahani, Abolfazl Rahimnejad, Hadis Karimipour
School of Engineering
University of Guelph
Guelph, Canada
{hkarimi, hmoham15, arahimne}@uoguelph.ca

*Abstract*— **Nowadays the emerging smart grid technology opens up the possibility of two-way communication between customers and energy utilities. Demand Response Management (DRM) offers the promise of saving money for commercial customers and households while helps utilities operate more efficiently. In this paper, an Incentive-based Demand Response Optimization (IDRO) model is proposed to efficiently schedule household appliances for minimum usage during peak hours. The proposed method is a multi-objective optimization technique based on Nonlinear Auto-Regressive Neural Network (NAR-NN) which considers energy provided by the utility and rooftop installed photovoltaic (PV) system. The proposed method is tested and verified using 300 case studies (household). Data analysis for a period of one year shows a noticeable improvement in power factor and customers' bill.**

## NOMENCLATURE

| | |
|---|---|
| $m$ | Total number of uninterruptible loads |
| $H$ | Time slot, 30 minutes |
| $P_C^t$ | Value of actual consumption at time $t$ |
| $P_O^t$ | The objective curve at time $t$ |
| $D_s$ | Customer's discomfort associated with a delay |
| $d_s$ | Customer discomfort associated with a shift |
| $x_n^s$ | Binary variables for device $s$ in time slot $n$ |
| $P_{sh_m}^H$ | Consumption of device $m$ in time slot $H$ |
| $B_H$ | Solar cell operating status (Binary) in time slot $H$ |
| $T_{App_i}$ | Operation time of device $i$ |
| $h_{App_i}$ | Operation time slots of the device $i$ |
| $P_{BSL}^H$ | Solar cell battery charge level in time slot $H$ |
| $T_B$ | The time required for the battery to be fully charged |
| $P_{All_m}^H$ | Total consumption of all loads in time slot $H$ |
| $D_c$ | Number of devices of type $k$ available for control |
| $P_{All}^H$ | Total consumption of shiftable devices in the time |
| $f_j^t$ | Number of fixed devices of type $j$ |
| $P_i$ | Power consumption of the device type $i$ |
| $P_P^t$ | Predicted load consumption at time $t$ |
| $\bar{P}_{off}$ | Average consumption during off-load |
| $L_{min}$ | Minimum usage required during off-peak hours |
| $P_{permitted}^{max}$ | Maximum consumption during peak hours |
| $[T_{peak_i}^{Start}, T_{peak_i}^{End}]$ | Start and end time interval for Peak hour $i$ |
| $CP_m^H$ | Number of preferred time slots for shifting determined by load $m$ in time slot $H$ |
| $s_i^t$ | Number of shiftable devices of type $i$ shifted to time $t$ |
| $s_k^t$ | Number of shiftable devices of type $k$ shifted away from time $t$ |

## I. INTRODUCTION

According to the U.S. Department of Energy, demand for electricity is expected to grow 30% by 2035 as a result of new consumption models (smart appliances, electric vehicles and whole house monitoring systems) [1]. Integration of the smart grid technology into the bulk power system provides the opportunity for two-way communication between the utility company and end-users through Demand Response Management (DRM) [2]. A detailed study of the potential impact of residential demand-side management on the cost and greenhouse gas emissions is presented in [3].

Although the future of DRM depends on automatic control of residential loads, the end-users play a significant role by shifting the use of appliances to the off-peak hours [4]. Different techniques are required to act as a bridge between the consumer and the utility for controlling the load demand during peak hours. However, in most techniques, incentive-based DR programs play a major role in improving grid operation and reliability as well as cost management [5].

Most of the researches on DRM are limited to high voltage levels, such as industrial loads [6]. There are few studies focused on the household sector [7]. In most of the proposed methods, customers' common welfare, utility costs, efficient operation, the cost of required communication infrastructure, and the probability of cyber-attack are neglected and generally focuses on only a single objective [8]. Some load control strategies for shedding household appliances [9] and several scheduling methods for mitigating residential power consumption [10]-[12] have been proposed in the past.

However, this study proposes an intelligent and flexible algorithm for DRM considering the combined source of energy, including electricity provided by the utility and rooftop installed residential photovoltaic (PV) system. The main contribution of this paper is the development of a scheduling algorithm (multi-objective optimization with NAR-NN) to minimize the electricity bill and customer's discomfort considering the operational dynamics of non-schedulable loads and electricity price variation. To avoid consumer's discomfort, a scheduling algorithm is applied, using historical data of the consumer's habits and PV generation forecasts.

## II. PROBLEM STATEMENT

The residential demand management problem is a non-linear programming problem aimed at minimizing power consumption and customers discomfort subject to some constraints.

### A. Power Consumption Minimization

Residential loads based on the power consumption pattern are categorized into two types:

- **Fixed loads:** whose power consumption and usage time cannot be modified (e.g., TV, refrigerator). It is assumed that the algorithm does not have any control on these loads.
- **Shiftable loads:** whose power consumption can be shifted to a different source or time slot to operate on its own power consumption pattern (e.g. air conditioner, washing machine, and dishwasher).

The goal is to schedules the consumption of each shiftable device to minimize the difference between the load consumption curve and the objective curve. Load shifting for power consumption minimization can be mathematically formulated as follows:

$$Minimize \quad J(t,p) = |P_C^t - P_O^t|^2 \qquad (1)$$

### B. Discomfort minimization

One of the main advantages of the proposed algorithm is to model load demand patterns based on the customers' lifestyle so that their discomfort can be minimized. The following minimization problem considering different weights $(w_s, k_s)$ is defined to minimize the customer discomfort:

$$Minimize \sum_{s \in S} w_s d_s + k_s D_s \qquad (2)$$

### C. Constraint

The minimization problem is subject to the following constraints:

1. The number of devices shifted cannot be a negative value.

$$s_i^t, f_j^t, s_k^t > 0 \quad \forall \, i,j,k \qquad (3)$$

2. The number of shiftable devices cannot be more than the number of devices available for control at each time step.

$$\sum_{k=1}^{\gamma} s_k^t \leq D_c \qquad (4)$$

3. Each shiftable devices is allowed to work in a specific time slot.

$$\sum_{n=T_{App}^{Start}}^{T_{App}^{Ens}-T_{App}+1} x_n^s = 1 \quad \forall \, s \in S \qquad (5)$$

## III. INCENTIVE-BASED DEMAND RESPONSE OPTIMIZATION (IDRO)

The proposed IDRO algorithm provides an opportunity to the consumers to voluntarily participate in the DRM. This algorithm considers the combined sources of energy provided by the grid and rooftop installed PV to implement DRM. As it was mentioned in previous section, loads are divided into fixed and shiftable ones. Each load can determine their permitted time used for shifting as the input of the algorithm. According to the schedules provided by the consumers, IDRO can decide whether their consumption is supplied through the grid or PV. This algorithm has several steps which are explained in this section.

### A. Methodology

Artificial neural network (ANN) is a well-known data processing algorithm to model non-linear systems. It works efficiently particularly when there are complex non-linear relationships between system input and output.

In this paper, a feed-forward neural network with Levenberg–Marquardt (LM) training algorithm is chosen for training the Nonlinear Auto-Regressive Neural Network (NAR-NN). According to the Fig. 1, the total consumption pattern for each hour of day during different months significantly varies.

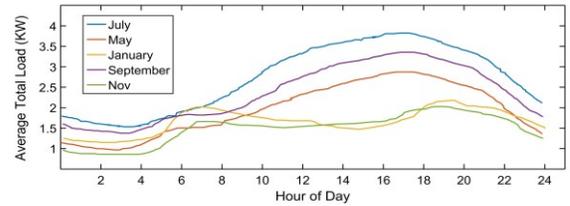

Fig. 1. Example of load consumption during different months of a year.

Therefore, in this network, different weight is assigned to historical data during a specific month/week of the year for each appliance. Historical data (network input) are used to predict the load consumption ahead of the time (network output). A hidden is a layer in between the input layers and output layers, where neurons receive a set of weighted inputs and generate an output using an activation function. The neural network architecture used in this work is given in Figure 2.

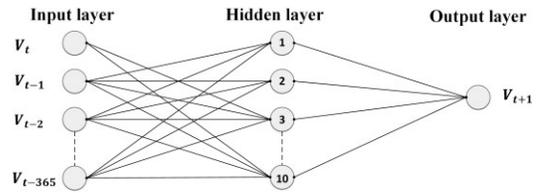

Fig. 2. NAR-NN basic scheme

The relationship between input and output of NA-RNN can be written as:

$$V(t) = h(V(t-1), V(t-2), \ldots, V(t-t_d)) + e(t) \quad (6)$$

where $e(t)$ is the error between real value and predicted value and $t_d$ is the size of data set.

### B. Objective Curve

Objective curve is inversely proportional to the electricity market prices. In offline mode, the objective curve is calculated based on the historical data and customers' preferences one day ahead. In online mode, the objective curve is updated every 30 minutes based on the real-time pricing and customer load consumption during the day. The objective is to shift the usage to 1) off-peak hours with a lower price, 2) energy generated by the solar cell, if available. An example of Real-Time Pricing (RTP) and solar cell energy generation is depicted in Fig. 3.

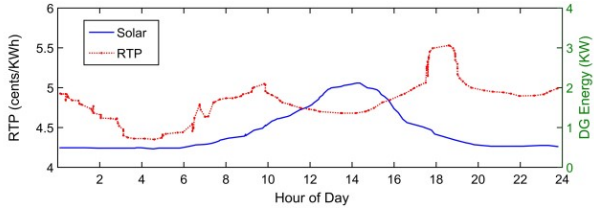

Fig. 3. Example of real-time pricing and solar cell energy generation during the day.

To have the lowest possible consumption during peak hours, the maximum permitted load is defined as follows:

$$P_O^t = \begin{cases} P_{permitted}^{max} & if \sum_{t \in T_o} \bar{P}_{off} < L_{min} \\ P_p^t & otherwise \end{cases} \quad (7)$$

To find the relationship between consumption at peak hours and off-peak hours RA is used in the proposed algorithm. The following relationship between the consumption in peak hours and off-peak hours is derived based on RA.

$$P_p^{max} = \sum_{i=0}^{y} \sum_{j=0}^{z} \alpha_{i,j} \bar{P}_{off,i} + \beta \quad (8)$$

where $\alpha, \beta, y$ and $z$ are constants calculated using RA.

### C. Load Consumption Optimization

The proposed algorithm determines the operation time of the shiftable loads and the type of the energy source (Solar cells or grid). This method is tested and verified using 300 case studies (household). For this purpose, each day is divided into 48 time slots of 30 minutes. The right-hand side matrix in (9), which is obtained by multiplying the power consumption matrix ($PCM_{m*48}$) and binary matrix for solar cell usage ($B_{48*1}$), shows whether in the time slot H the shiftable loads is supplied by solar cell or grid. It should be noted $B_H$ is a binary value which shows it solar cell in time slot H is in use or not (10).

$$\begin{bmatrix} P_{sh_1}^1 & \cdots & P_{sh_1}^{48} \\ \vdots & \ddots & \vdots \\ P_{sh_m}^1 & \cdots & P_{sh_m}^{48} \end{bmatrix} \times \begin{bmatrix} B_1 \\ \vdots \\ B_{48} \end{bmatrix} = \begin{bmatrix} \sum_{H=1}^{48} P_{sh_1}^H B_H \\ \vdots \\ \sum_{H=1}^{48} P_{sh_m}^H B_H \end{bmatrix} \quad (9)$$

$$B_H = \begin{cases} 1 & if: T_B - T_{Peak}^{Start} > Max\{T_{App}\}, P_{BSL}^H > \sum_{i=1}^{m} P_{sh_i}^H \\ 0 & else \end{cases} \quad (10)$$

The maximum acceptable shift in each time slot determined by the consumers can be written as:

$$CP = \begin{bmatrix} CP_1^1 & \cdots & CP_1^{48} \\ \vdots & \ddots & \vdots \\ CP_m^1 & \cdots & CP_m^{48} \end{bmatrix} \quad (11)$$

Then, considering costumer preferences matrix (11) and operation time slots of the device $i$ ($h_{App_i}$) applying them in (9) the following equation is derived:

$$\begin{bmatrix} P_{sh_1}^1 & \cdots & P_{sh_1}^{48} \\ \vdots & \ddots & \vdots \\ P_{sh_m}^1 & \cdots & P_{sh_m}^{48} \end{bmatrix} \times \begin{bmatrix} B_1 \\ \vdots \\ B_{48} \end{bmatrix} = \begin{bmatrix} \sum_{H=1}^{48} P_{sh_1}^{H+CP_1^h - h_{App_1}} B_H \\ \vdots \\ \sum_{H=1}^{48} P_{sh_m}^{H+CP_m^h - h_{App_m}} B_H \end{bmatrix} \quad (12)$$

It should be noted in (12):

$$H + CP_m^h - h_{App_m} > 30 \; minutes \quad (13)$$

Eventually, for the whole consumption in each time slot we can write:

$$\begin{bmatrix} P_{All}^1 \\ \vdots \\ P_{All}^{48} \end{bmatrix} = \begin{bmatrix} \sum_{i=1}^{m} P_i^1 B_1 \\ \vdots \\ \sum_{i=1}^{m} P_i^{48} B_{48} \end{bmatrix} \quad (14)$$

Fig. 4 shows the flowchart of the proposed IDRO.

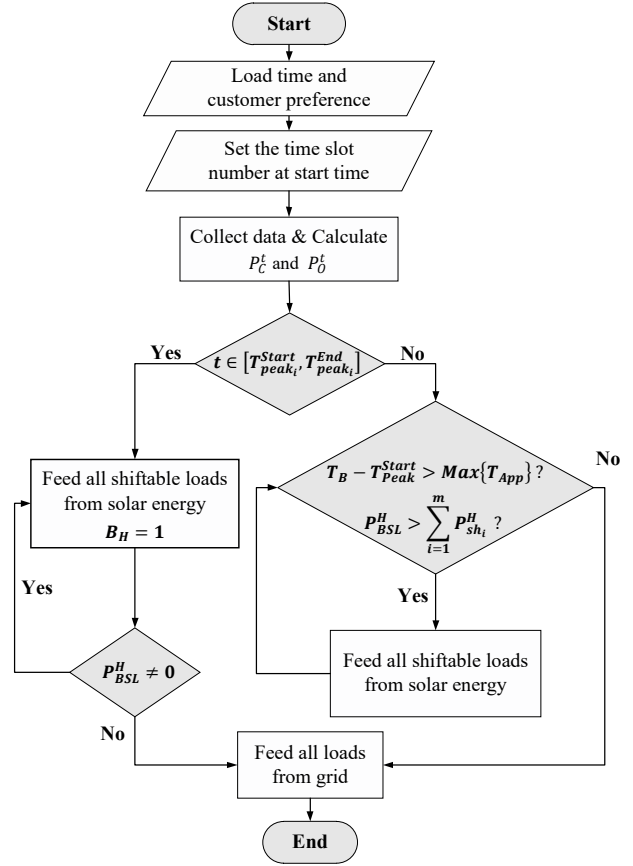

Fig. 4. Overall flowchart of the proposed IDRO method

## IV. CASE STUDIES AND RESULTS

The average results for 300 household consumers are presented in this section. MATLAB® software is used for data analysis. Fig. 5 shows the proposed energy management platform. In this study, air conditioner, dishwasher, laundry machine, iron, etc. are considered as shiftable loads and a refrigerator and one lamp are included as fixed loads.

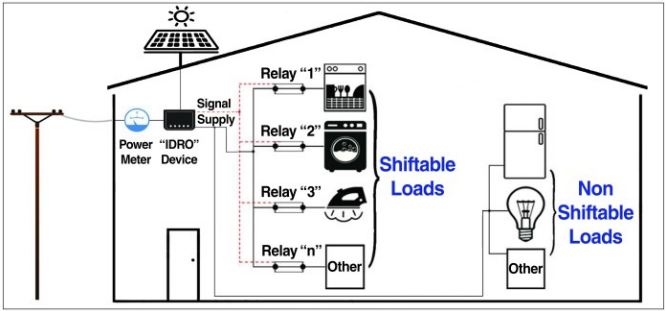

Fig. 5. IDRO energy management platform.

### A. Data acquisition

The input data for the period of one year is collected every two months from Hamedan province Electricity Distribution Company. In fact, the domestic consumption data of 300 consumers along with their hourly generation of a 1 Kw of a solar cell is used as input data.

### B. Training, Test, and Validation

For the training purpose, Levenbrg-Marquardt is applied to using 6132 data out of 8760 (70 percent) and for each test and validation purpose, a set of 1314 (15 percent) data is chosen randomly out of the whole data. The data is processed to predict the load curve and PV generation rate for each day using three layers (input, output and hidden, containing 24, 1 and 10 neurons, respectively). Different number are tested and the best performance is obtained with 10 neurons. The training procedure automatically stops when there is no significant change in Mean Squared Error (MSE). Fig. 6 demonstrates the training procedure accomplished by Levenberg–Marquardt backpropagation algorithm converged after 12 epochs. As it can be observed, there is no increase after (stability) and before (overshoot) convergence. Moreover, in the test and validation procedure, errors decrease until the Epoch 12.

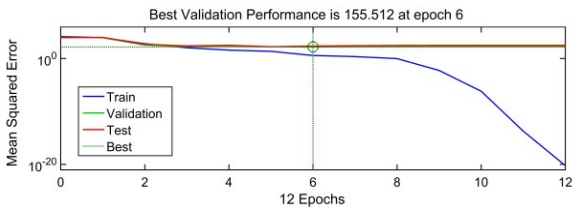

Fig. 6. Training performance graph.

Fig. 7 displays the error autocorrelation index. For a perfect prediction model, there should be at most one nonzero value of the autocorrelation function, and it should occur at zero lag. This metric indicate that there is no correlation between data. It also determines the presence of correlation between the values of variables which are based on associated aspects. As can be seen in the Fig. 7, in the proposed model, the correlations, except for the one at zero lag, fall approximately within the 95% confidence limits which confirms the accuracy of the model.

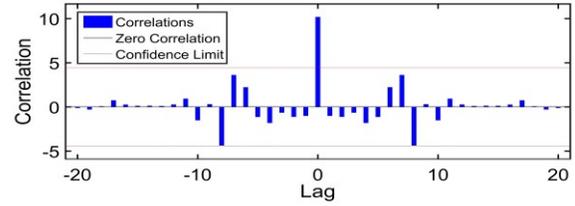

Fig. 7. Error Autocorrelation

A typical day-ahead prediction load consumption and PV generation are carried out using NAR-NN and the results are depicted in Fig. 8. As can be seen, the predicted load consumption and the predicted PV generation track the real ones with an acceptable precision. It should be mentioned that the PV generation prediction for the considered region provided in [13] verifies the obtained result.

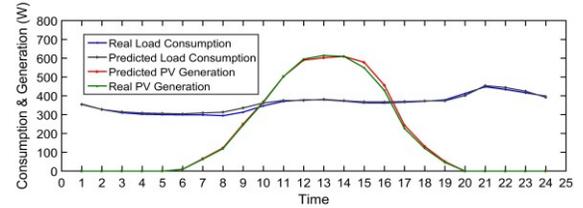

Fig. 8. Predicted Load Consumption and PV generation

### C. Load reduction during peak hours

The load profile for a household, before and after IDRO implementation and the PV generation are depicted in Fig. 9. As shown in the figure, using the proposed method, the load profile is leveled off and the consumption is distributed during the day. In addition, the average power consumption, for all case studies in each period is depicted in Fig. 10. As shown, the consumption rate reduced by almost 19%.

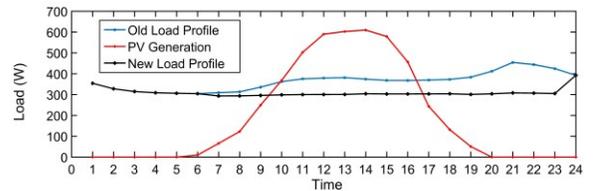

Fig. 9. Load profiles and PV generation

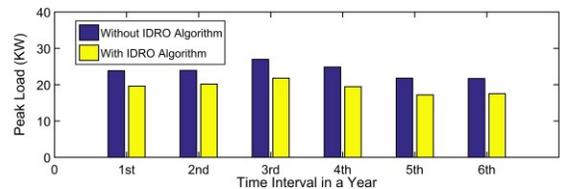

Fig. 10. Average energy consumption in a year.

### D. Load factor improvement

Fig. 11 shows the average power factor improvement using the proposed IDRO method. It clearly indicates that this method improves the power factor by %11-%17 during the different period of the year. It should be noted that if the proposed method is implemented properly and on a large scale, it can save a huge amount of energy.

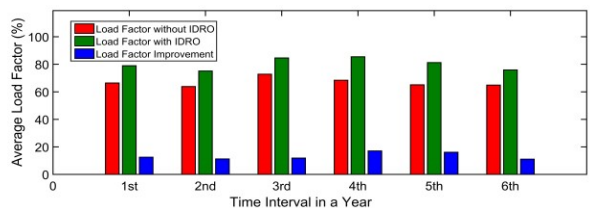

Fig. 11. Overall power factor improvement.

*E. Bill reduction*

The customers' bills for all case studies during six time intervals in a year are calculated. Fig. 12 shows the average reduction in the customers' bills for all case studies. The results show that this average amount is reduced by 56%. The tariff rate used in this work is also reported on [14].

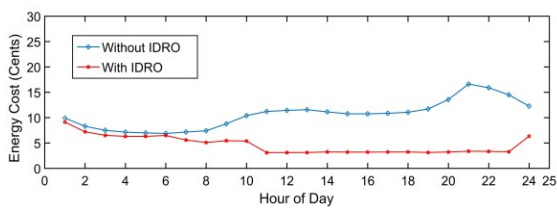

Fig. 12. Average cost reduction using the IDRO algorithm for all case studies.

## V. CONCLUSION

In this paper, a new residential load management algorithm based on multi-objective optimization combined with NAR-NN is proposed. One of the main advantages of the proposed algorithm is its flexibility which models load demand patterns based on the lifestyles of the customers. The algorithm does not attempt to reduce the consumption but tries to provide an optimal pattern for the peak hours considering the consumption behavior for each specific consumer. The results show an average of 19% consumption reduction on peak hours and up to 17% load factor improvement. It consequently reduced the customers' bills by almost 56%.

The proposed method can be easily implemented in a small electronic device at a minimum cost. Moreover, in contrast with the existing method which requires major changes in different home appliances, the proposed device can be separately added to the appliances at minimum cost and with the same accuracy. It is also possible to adapt the device's scheduling with new conditions if the peak load hours are changed.